# Derivation of Relations between Scaling Exponents and Standard Deviation Ratios

Yanguang Chen

(Department of Geography, College of Urban and Environmental Sciences, Peking University, 100871, Beijing, China. Email: chenyg@pku.edu.cn)

**Abstract**: The law of allometric growth is one of basic rules for understanding urban evolution. The general form of this law is allometric scaling law. However, the deep meaning and underlying rationale of the scaling exponents remain to be brought to light. In this paper, the theories of linear algebra and regression analysis are employed to reveal the mathematical and statistic essence of allometric scaling exponents. Suppose that the geometric measure relations between a set of elements in an urban system follow the allometric growth law. An allometric scaling exponent is proved to equal in theory to the ratio of the standard deviation of one logarithmic measure to the standard deviation of another logarithmic measure. In empirical analyses based on observational data, the scaling exponent is equal to the product between the standard deviation ratio and the corresponding Pearson correlation coefficient. The mathematical derivation results can be verified by empirical analysis: the scaling exponent values based on the standard deviation ratios are completely identical to those based on the conventional method. This finding can be generalized to city fractals and city size distribution to explain fractal dimensions of urban space and Zipf scaling exponent of urban hierarchy. A conclusion can be reached that scaling exponents reflect the ratios of characteristic lengths. This study may be helpful for comprehending scaling from a new perspective and the connections and distinctions between scaling and characteristic scales.

# 1 Introduction

Allometric scaling analysis is one of scale-free analysis methods, which can be used to study urban growth and form as well as urban hierarchy. In scientific research, mathematical modeling and quantitative analysis fall into two types: one is based on characteristic scales, and the other is based on scaling (Mandelbrot, 1982; Feder, 1988; Takayasu, 1990). If a system bears characteristic lengths such as determinate radius, perimeter, mean, standard deviation, and eigenvalues, it belong to the systems with characteristic scales, can be modeled and analyzed by using conventional mathematical methods. In contrast, if a system bears no characteristic length, it belongs to scale-free systems, and cannot be described with conventional mathematical methods. Scale-free systems are all complex systems. Scaling analysis is one of powerful tools to research complex systems. Allometric scaling relations are in essence a fractal measure relation. An allometric scaling exponent (ASE) proved to the ratio of two fractal dimensions and the ratio of two relative growth rates (Chen, 2008; Chen and Jiang, 2009). Fractal dimensions correspond to space, and relative growth rates correspond to time. Using the allometric scaling exponents, we can explore the spatio-temporal processes and patterns of urban evolution. Thus, an analytical process based on multi-scaling allometry was put forward to evaluate city development and interpret spatial heterogeneity of city distributions (Chen, 2017).

The mathematical and statistic nature of the allometric scaling exponents is not yet clear. In previous studies, an empirical relationship between allometric scaling exponents and standard deviations was disclosed. The calculated value of an allometric scaling exponent equals the product between the ratio of the standard deviation of one logarithmic measure to the standard deviation of another correlated logarithmic measure of an urban system and the corresponding Pearson correlation coefficient. However, three problems remain to be solved. First, what is the theoretical relation between allometric scaling exponents and standard deviations? Second, how to prove the theoretical and empirical relationships? Third, whether or not the parameter relations can be generalized to fractal dimension and rank-size scaling exponent? This paper is devoted to answering these questions by theoretical derivation and empirical evidence. The rest of this paper is organized as follows. In Section 2, a complete mathematical derivation process will be presented. The derived results construct the underlying rationale of the methodology of multi-scaling allometric analysis.

In Section 3, a simple case study will be made to show testify the theoretical derivation results for the mathematical essence of allometric scaling exponents. In Section 4, several questions will be discussed. Finally, the discussion will be concluded by summarizing the main points of this work.

## 2 Mathematical models

### 2.1 Theoretical models and parameter relations

Allometric scaling relations can be regarded as type of geometric measure relations. A geometric measure relation is a proportional relationship between two measures. A measure (e.g., urban area) is proportional to another measure (e.g., city population size) if and only if the two measures share the same dimension. The allometric growth indicates that the ratio of the relative growth rate of one measure to the relative growth rate of another correlated measure is a constant not equal to 1. The constant ratio is what is called allometric scaling exponent. The exponent equaling 1 implies isometric growth, which can be treated as a special case of allometric growth (Lee, 1989). Taking cities as an example, we can reveal the mathematical and statistic essence of the scaling exponents. Suppose that there are $n$ cities in a geographical region. Starting from general system theory, we can derive apriori relations of allometric growth as follows (Bertalanffy, 1968)

$$Q_i(t) = \beta_{ij} Q_j(t)^{\alpha_{ij}}, \quad Q_j(t) = \beta_{ji} Q_i(t)^{\alpha_{ji}}, \tag{1}$$

where time, $Q_i(t)$ and $Q_j(t)$ denote two measures of cities $i$ and $j$ at time $t$, $\alpha_{ij}$ and $\alpha_{ji}$ refer to two scaling exponents, and $\beta_{ij}$ and $\beta_{ji}$ to two proportionality coefficients ($i, j=1,2,3,\ldots, n$). Take natural logarithms of equation (1) yields a pair of linear relations such as

$$\ln Q_i(t) = \ln \beta_{ij} + \alpha_{ij} \ln Q_j(t), \quad \ln Q_j(t) = \ln \beta_{ji} + \alpha_{ji} \ln Q_i(t). \tag{2}$$

Define the means of the logarithmic time series as below:

$$\mu_i = \frac{1}{T} \sum_{t=1}^{T} \ln Q_i(t), \mu_j = \frac{1}{T} \sum_{t=1}^{T} \ln Q_j(t), \tag{3}$$

where $\mu_i$ and $\mu_j$ refer to arithmetic means, $t=1,2,3,\ldots, T$, and $T$ represents the length of sample path extracting from time series. Correspondingly, the population standard deviations are

$$\sigma_i = [\frac{1}{T} \sum_{t=1}^{T} (\ln Q_i(t) - \mu_i)^2]^{1/2}, \sigma_j = [\frac{1}{T} \sum_{t=1}^{T} (\ln Q_j(t) - \mu_j)^2]^{1/2}, \tag{4}$$

where $\sigma_i$ and $\sigma_j$ refer to two standard deviations. According to the theory of linear regression, the slopes, i.e., the regression coefficients, of equation (2) can be given by

$$\alpha_{ij} = \frac{\sum_{t=1}^{T}(\ln Q_j(t)-\mu_j)(\ln Q_i(t)-\mu_i)}{\sum_{t=1}^{T}(\ln Q_j(t)-\mu_j)^2}, \quad \alpha_{ji} = \frac{\sum_{t=1}^{T}(\ln Q_i(t)-\mu_i)(\ln Q_j(t)-\mu_j)}{\sum_{t=1}^{T}(\ln Q_i(t)-\mu_i)^2}. \tag{5}$$

The two expressions in equation (1) are inverse functions to each other. Therefore, the two scaling exponents are reciprocal to one another. That is,

$$\alpha_{ij}\alpha_{ji} = 1. \tag{6}$$

Divide two sides of equation (6) by $\alpha_{ij}^2$ produces

$$\frac{\alpha_{ji}}{\alpha_{ij}} = \frac{1}{\alpha_{ij}^2}. \tag{7}$$

Substituting equation (5) into equation (7) yields

$$\alpha_{ij} = \sqrt{\frac{\alpha_{ij}}{\alpha_{ji}}} = \frac{\sqrt{\sum_{t=1}^{T}(\ln Q_i(t)-\mu_i)^2}}{\sqrt{\sum_{t=1}^{T}(\ln Q_j(t)-\mu_j)^2}} = \frac{\sigma_i}{\sigma_j}. \tag{8}$$

This suggests that a scaling exponent, $\alpha_{ji}$, is a ratio of two standard deviations, $\sigma_i/\sigma_j$. Similarly, we have $\alpha_{ji}=\sigma_j/\sigma_i$. The relation, equation (8), was empirically found in previous work (Chen, 2017), but not mathematical demonstrated.

The allometric scaling exponent bears other mathematical and physical meanings. On the one hand, as indicated above, it represents a ratio of relative rates of growth in time. Differentiating equation (2) with respect to time $t$ yields

$$\alpha_{ij} = \frac{d\ln Q_i(t)}{d\ln Q_j(t)} = \frac{dQ_i(t)/Q_i(t)dt}{dQ_j(t)/Q_j(t)dt} = \frac{r_i}{r_j}, \quad \alpha_{ji} = \frac{d\ln Q_j(t)}{d\ln Q_i(t)} = \frac{dQ_j(t)/Q_j(t)dt}{dQ_i(t)/Q_i(t)dt} = \frac{r_j}{r_i}, \tag{9}$$

in which $r_i$ and $r_j$ denotes relative growth rates of $Q_i(t)$ and $Q_j(t)$, that is,

$$r_i = \frac{d(\ln Q_i(t))}{dt} = \frac{dQ_i(t)}{Q_i dt}, \quad r_j = \frac{d(\ln Q_j(t))}{dt} = \frac{dQ_j(t)}{Q_j dt}. \tag{10}$$

On the other hand, it is a ratio of two correlated fractal dimensions for space. According to the principle of dimensional consistency (Lee, 1989; Mandelbrot, 1982; Takayasu, 1990), equation (1) can be re-expressed as

$$Q_i(t) = \beta_{ij} Q_j(t)^{D_i/D_j}, \quad Q_j(t) = \beta_{ji} Q_i(t)^{D_j/D_i}. \tag{11}$$

Thus we have a useful parameter equation as follows (Chen, 2017)

$$\alpha_{ij} = \frac{r_i}{r_j} = \frac{D_i}{D_j} = \frac{\sigma_i}{\sigma_j}. \tag{12}$$

Based on equation (12), a reciprocal matrix of allometric scaling exponents can be constructed as below (Chen, 2008; Chen and Jiang, 2009)

$$\mathbf{M} = \begin{bmatrix} 1 & r_1/r_2 & \cdots & r_1/r_n \\ r_2/r_1 & 1 & \cdots & r_2/r_n \\ \vdots & \vdots & \ddots & \vdots \\ r_n/r_1 & r_n/r_2 & \cdots & 1 \end{bmatrix} = \begin{bmatrix} 1 & D_1/D_2 & \cdots & D_1/D_n \\ D_2/D_1 & 1 & \cdots & D_2/D_n \\ \vdots & \vdots & \ddots & \vdots \\ D_n/D_1 & D_n/D_2 & \cdots & 1 \end{bmatrix}. \tag{13}$$

In theory, equation (13) can be replaced by

$$\mathbf{M} = \begin{bmatrix} 1 & \sigma_1/\sigma_2 & \cdots & \sigma_1/\sigma_n \\ \sigma_2/\sigma_1 & 1 & \cdots & \sigma_2/\sigma_n \\ \vdots & \vdots & \ddots & \vdots \\ \sigma_n/\sigma_1 & \sigma_n/\sigma_2 & \cdots & 1 \end{bmatrix} = \left[ \sigma_i/\sigma_j \right]. \tag{14}$$

which have be empirically supported by observed data (Chen, 2017). Consequently, the spatial and temporal structure behind the urban allometry has been converted into a type of probability structure.

**2.2 Matrix scaling equation of standard deviation ratios**

The allometric scaling exponent has been proved to be a ratio of one standard deviation to another standard deviation. This proof advances our understanding of allometric scaling exponent. Based on this result, a new algorithm can be proposed to estimate the values of scaling exponents. Define two vectors of population standard deviations as follows

$$\mathbf{\Sigma}^{\mathrm{T}} = [\sigma_1 \quad \sigma_2 \quad \cdots \quad \sigma_2], \mathbf{\Sigma}^{*\mathrm{T}} = [1/\sigma_1 \quad 1/\sigma_2 \quad \cdots \quad 1/\sigma_2]. \tag{15}$$

Thus the outer product of the two vectors is

$$\mathbf{\Sigma}\mathbf{\Sigma}^{*\mathrm{T}} = \begin{bmatrix} \sigma_1 \\ \sigma_2 \\ \vdots \\ \sigma_n \end{bmatrix} [1/\sigma_1 \quad 1/\sigma_2 \quad \cdots \quad 1/\sigma_n] = \begin{bmatrix} 1 & \sigma_1/\sigma_2 & \cdots & \sigma_1/\sigma_n \\ \sigma_2/\sigma_1 & 1 & \cdots & \sigma_2/\sigma_n \\ \vdots & \vdots & \ddots & \vdots \\ \sigma_n/\sigma_1 & \sigma_n/\sigma_2 & \cdots & 1 \end{bmatrix} = \mathbf{M}. \tag{16}$$

The inner product of the two vectors is equal to the sample size, $n$, that is

$$\boldsymbol{\Sigma}^{*\mathrm{T}}\boldsymbol{\Sigma}=[1/\sigma_1 \quad 1/\sigma_2 \quad \cdots \quad 1/\sigma_n]\begin{bmatrix}\sigma_1\\\sigma_2\\\vdots\\\sigma_n\end{bmatrix}=[\sigma_1 \quad \sigma_2 \quad \cdots \quad \sigma_n]\begin{bmatrix}1/\sigma_1\\1/\sigma_2\\\vdots\\1/\sigma_n\end{bmatrix}=\boldsymbol{\Sigma}^{\mathrm{T}}\boldsymbol{\Sigma}^{*}=n. \quad (17)$$

Apparently, we have

$$\begin{bmatrix}1 & \sigma_1/\sigma_2 & \cdots & \sigma_1/\sigma_n\\\sigma_2/\sigma_1 & 1 & \cdots & \sigma_2/\sigma_n\\\vdots & \vdots & \ddots & \vdots\\\sigma_n/\sigma_1 & \sigma_n/\sigma_2 & \cdots & 1\end{bmatrix}\begin{bmatrix}\sigma_1\\\sigma_2\\\vdots\\\sigma_n\end{bmatrix}=n\begin{bmatrix}\sigma_1\\\sigma_2\\\vdots\\\sigma_n\end{bmatrix}, \quad (18)$$

which can be simply expressed as

$$\mathbf{M}\boldsymbol{\Sigma}=n\boldsymbol{\Sigma}. \quad (19)$$

This implies that the standard deviation vector is the eigenvector of the outer product of the standard deviation, and the inner product of the standard deviation is just the corresponding maximum eigenvalue. Equation (19) can be expanded as follows

$$\begin{bmatrix}\sigma_1\\\sigma_2\\\vdots\\\sigma_n\end{bmatrix}[1/\sigma_1 \quad 1/\sigma_2 \quad \cdots \quad 1/\sigma_n]\begin{bmatrix}\sigma_1\\\sigma_2\\\vdots\\\sigma_n\end{bmatrix}=[1/\sigma_1 \quad 1/\sigma_2 \quad \cdots \quad 1/\sigma_n]\begin{bmatrix}\sigma_1\\\sigma_2\\\vdots\\\sigma_n\end{bmatrix}\begin{bmatrix}\sigma_1\\\sigma_2\\\vdots\\\sigma_n\end{bmatrix}, \quad (20)$$

in which the standard deviations can be replaced by fractal dimension values and relative growth rates. Using equation (18), or equation (19), or Equation (20), we can make multiple allometric scaling analysis of cities in a geographical region or urban elements in a city. Before applying the model to actual problems in the real world, we should improve it in terms of mathematical and statistical theories.

## 2.3 Algorithm based on standard deviation ratios

The essence of scaling exponent lies in the ratios of two standard deviations, and this discovery suggests a new algorithm for estimating allometric scaling indexes in a multi-scaling allometric framework. From equations (5) and (6) it follows

$$\alpha_{ij}\alpha_{ji}=\frac{(\sum_{t=1}^{T}(\ln Q_i^{(t)}-\mu_i)(\ln Q_j^{(t)}-\mu_j))^2}{\sum_{t=1}^{T}(\ln Q_j^{(t)}-\mu_j)^2\sum_{t=1}^{T}(\ln Q_j^{(t)}-\mu_j)^2}=R_{ij}^2=1, \quad (21)$$

where $R_{ij}$ denotes Pearson correlation coefficient, and $R_{ij}=R_{ji}$. However, where observational data with random disturbance is concerned, $R_{ij}^2<1$. Thus we have

$$\alpha_{ij}^*\alpha_{ji}^* = R_{ij}^2 = R_{ji}^2 < 1, \tag{22}$$

where the parameters with asterisk "*" means calculated values rather than theoretical values. In practice, the population standard deviation $\sigma$ should be substituted by sample standard deviation $s$, and equation (18) should be replaced by

$$\mathbf{MS} = \begin{bmatrix} 1 & s_1/s_2 & \cdots & s_1/s_n \\ s_2/s_1 & 1 & \cdots & s_2/s_n \\ \vdots & \vdots & \ddots & \vdots \\ s_n/s_1 & s_n/s_2 & \cdots & 1 \end{bmatrix} \begin{bmatrix} s_1 \\ s_2 \\ \vdots \\ s_n \end{bmatrix} = n \begin{bmatrix} s_1 \\ s_2 \\ \vdots \\ s_n \end{bmatrix} = n\mathbf{S}, \tag{23}$$

where $\mathbf{S}=[s_1\ s_2\ \ldots\ s_n]^T$ refers to the sample standard deviation vector, $\mathbf{M}$ denotes the sample standard deviation matrix rather than the population standard deviation matrix. For simplicity, the symbol $\mathbf{M}$ does not change. The sample standard deviations can be given by

$$s_i = [\frac{1}{T-1}\sum_{t=1}^{T}(\ln Q_i(t)-\mu_i)^2]^{1/2}, s_j = [\frac{1}{T-1}\sum_{t=1}^{T}(\ln Q_j(t)-\mu_j)^2]^{1/2}. \tag{24}$$

In light of equation (22), we have

$$\alpha_{ij}^*\alpha_{ji}^* = \alpha_{ij}^*\frac{s_j}{s_i}\frac{s_i}{s_j}\alpha_{ji}^* = R_{ij}R_{ji}, \tag{25}$$

which can be decomposed into

$$\alpha_{ij}^* = R_{ij}\frac{s_i}{s_j}, \alpha_{ji}^* = R_{ji}\frac{s_j}{s_i}. \tag{26}$$

This implies that the empirical scaling exponent based on standard deviation ratios should be adjusted by correlation coefficient.

It is easy to calculate standard deviation, and then obtain the outer product between standard deviations and reciprocals of standard deviations. The logarithmic values of the time series can be standardized by the following formulae

$$y_i = \frac{\ln Q_i(t)-\mu_i}{s_i}, y_j = \frac{\ln Q_j(t)-\mu_j}{s_j}, \tag{27}$$

where $y_i$ and $y_j$ represent the standardized results of $\ln Q_i(t)$ and $\ln Q_j(t)$, namely, Z-scores. Thus, we have matrix $\mathbf{Y}=[\mathbf{y}_1, \mathbf{y}_2,\ldots,\mathbf{y}_n]$, where the standardize vector $\mathbf{y}_j=[y_{1j}, y_{2j}, \ldots, y_{Tj}]^T$. Then, the correlation coefficient matrix can be calculated by the following equation

$$\mathbf{V} = \frac{1}{n-1}\mathbf{Y}^{\mathrm{T}}\mathbf{Y} = \left[R_{ij}\right] = \begin{bmatrix} 1 & R_{12} & \cdots & R_{1n} \\ R_{21} & 1 & \cdots & R_{2n} \\ \vdots & \vdots & \ddots & \vdots \\ R_{n1} & R_{n2} & \cdots & 1 \end{bmatrix}. \tag{28}$$

Making array multiplication between **R** and **M** yields an adjusted sample standard deviation matrix

$$\mathbf{M}^* = \begin{bmatrix} 1 & R_{12}s_1/s_2 & \cdots & R_{1n}s_1/s_n \\ R_{21}s_2/s_1 & 1 & \cdots & R_{2n}s_2/s_n \\ \vdots & \vdots & \ddots & \vdots \\ R_{n1}s_n/s_1 & R_{n2}s_n/s_2 & \cdots & 1 \end{bmatrix} = \begin{bmatrix} 1 & \alpha_{12}^* & \cdots & \alpha_{1n}^* \\ \alpha_{21}^* & 1 & \cdots & \alpha_{2n}^* \\ \vdots & \vdots & \ddots & \vdots \\ \alpha_{n1}^* & \alpha_{n2}^* & \cdots & 1 \end{bmatrix}. \tag{29}$$

Thus, equation (23) should be substituted with

$$\mathbf{M}^*\mathbf{S}^* = \begin{bmatrix} 1 & R_{12}s_1/s_2 & \cdots & R_{1n}s_1/s_n \\ R_{21}s_2/s_1 & 1 & \cdots & R_{2n}s_2/s_n \\ \vdots & \vdots & \ddots & \vdots \\ R_{n1}s_n/s_1 & R_{n2}s_n/s_2 & \cdots & 1 \end{bmatrix} \begin{bmatrix} s_1^* \\ s_2^* \\ \vdots \\ s_n^* \end{bmatrix} = \lambda_{\max} \begin{bmatrix} s_1^* \\ s_2^* \\ \vdots \\ s_n^* \end{bmatrix} = \lambda_{\max}\mathbf{S}^*, \tag{30}$$

where $\mathbf{S}^* = [s_1^*\ s_2^*\ \ldots\ s_n^*]^{\mathrm{T}}$ denotes the adjusted vector of sample standard deviation, $\mathbf{M}^*$ refers to the adjusted matrix of sample standard deviations.

## 3 Empirical analysis

### 3.1 Study area, datasets, and algorithms

The multi-scaling allometric analysis can be used to evaluate city development and generate a rank of the relative growth potential for cities or elements of a city. This has been demonstrated for a long time (Chen, 2008; Chen, 2017; Chen and Jiang, 2009; Long and Chen, 2019). This study is devoted to showing the mathematical essence of allometric scaling exponents and verifying the related mathematical derivation results. As a simple example, the four municipalities directly under the Central Government of China, Beijing (BJ), Tianjin (TJ), Shanghai (SH), and Chongqing (CQ), are employed to make a case study (Figure 1). This example can be used to confirm the abovementioned mathematical deduction. The basic measurement is gross domestic product (GDP) (Table 1). To find a solution to a model, we need effective algorithms. Next, two algorithms are adopted to calculate the allometric scaling exponent matrix. One algorithm is based on the above mathematical derivation results, the other is the conventional method, i.e., the double logarithmic linear regression based on ordinary least squares (OLS) used in the past (Chen, 2008; Chen, 2017). If the results from the two methods are the same with one another, it shows that the above mathematical reasoning process is correct and acceptable.

The first approach is matrix operation based on the ratio of the standard deviation of one logarithmic variable to the standard deviation of another logarithmic variable. The principle was made clear in subsection 2.3. This method involves the inter product and outer product between standard deviation vector and vector of standard deviation reciprocals as well as array multiplication of the correlation coefficient matrix and scaling exponent matrix. The key lies in the ratios of the standard deviations of the logarithmic variables, so the algorithm was briefly termed *standard deviation ratios* (SDR) method. The procedure is summarized as follows. **Step 1**: take the logarithm of the observational variables. The formula is $x_j = \ln Q_j(t)$, where $t=1,2,\ldots,T$ refers to time, and $T$ to the length of sample path ($j=1,2,\ldots,n$). The results form a vector $\mathbf{x}_j=[x_{1j}\ x_{2j}\ \ldots\ x_{Tj}]^T$, which makes a matrix of logarithmic variables such as $\mathbf{X}=[\mathbf{x}_1\ \mathbf{x}_2\ \ldots\ \mathbf{x}_n]$. **Step 2**: standardize the logarithmic variables. The formula is equations (27), which can be rewritten as $\mathbf{y}_j=(\mathbf{x}_j-\mu_j)/s_j$, where $\mu_j$ refers to the mean of $x_j$, and $s_j$ is the corresponding sample standard deviation. The results form a matrix of standardized logarithmic variables $\mathbf{Y}=[\mathbf{y}_1\ \mathbf{y}_2\ \ldots\ \mathbf{y}_n]$. **Step 3**: compute the matrix of standard deviation ratios. The method is to make use of outer product between the vector of standard deviations and the vector of standard deviation reciprocals. The formula is equations (16). It is easy to work out standard deviations using electronic spreadsheet or statistical software. **Step 4**: calculate the logarithmic linear correlation coefficients. We can obtain the Pearson correlation coefficients by means of matrix multiplication. Based on the *sample standard deviation* (SSD), the formula is equations (28). If we use the *population standard deviation* (PSD) to standardize the logarithmic random variables for theoretical analyses, the formula should be substituted by $\mathbf{V}=\mathbf{Y}^T\mathbf{Y}/n$. **Step 5**: compute the scaling exponent matrix. The matrix can be gotten by array multiplication of correlation coefficient matrix and the matrix of standard deviation ratios. The formula is equations (29), which is quasi-reciprocal matrix rather than a strict reciprocal matrix. **Step 6**: calculate the eigenvector and the maximum eigenvalue of the scaling exponent matrix. Normalizing the eigenvector yields the scaling indexes of allometric growth. The allometric scaling indexes can be used to make spatial analysis of urban growth, and the maximum eigenvalues can be employed to make model test.

**Table 1 The gross regional product (GRP) of four Chinese municipalities (1998-2018)**

Unit: 100 millions *yuan*

| Year | Beijing | Tianjin | Shanghai | Chongqing |
|---|---|---|---|---|
| 1998 | 2816.818 | 1463.445 | 3994.190 | 1630.671 |
| 1999 | 3045.308 | 1587.933 | 4369.718 | 1688.230 |
| 2000 | 3471.477 | 1795.232 | 4928.734 | 1813.309 |
| 2001 | 3985.302 | 2015.059 | 5361.584 | 1996.347 |
| 2002 | 4499.366 | 2246.187 | 5857.495 | 2249.095 |
| 2003 | 5219.790 | 2692.410 | 6830.032 | 2568.725 |
| 2004 | 6296.744 | 3248.995 | 8236.599 | 3043.394 |
| 2005 | 6969.520 | 3905.640 | 9247.660 | 3467.720 |
| 2006 | 8117.780 | 4462.740 | 10572.240 | 3907.230 |
| 2007 | 9846.810 | 5252.760 | 12494.010 | 4676.130 |
| 2008 | 11115.000 | 6719.010 | 14069.860 | 5793.660 |
| 2009 | 12153.030 | 7521.850 | 15046.450 | 6530.010 |
| 2010 | 14113.580 | 9224.460 | 17165.980 | 7925.580 |
| 2011 | 16251.930 | 11307.280 | 19195.690 | 10011.370 |
| 2012 | 17879.400 | 12893.880 | 20181.720 | 11409.600 |
| 2013 | 19800.810 | 14442.010 | 21818.150 | 12783.260 |
| 2014 | 21330.830 | 15726.930 | 23567.700 | 14262.600 |
| 2015 | 23014.590 | 16538.190 | 25123.450 | 15717.270 |
| 2016 | 25669.130 | 17885.390 | 28178.650 | 17740.590 |
| 2017 | 28014.940 | 18549.190 | 30632.990 | 19424.730 |
| 2018 | 30319.980 | 18809.640 | 32679.870 | 20363.190 |
| Mean | 13044.3874 | 8489.9158 | 15216.7986 | 8047.7481 |
| Stdev | 8844.3701 | 6371.8483 | 9180.3043 | 6374.8384 |

**Source**: National Bureau of Statistics of China, available from: http://www.stats.gov.cn/tjsj/ndsj/.

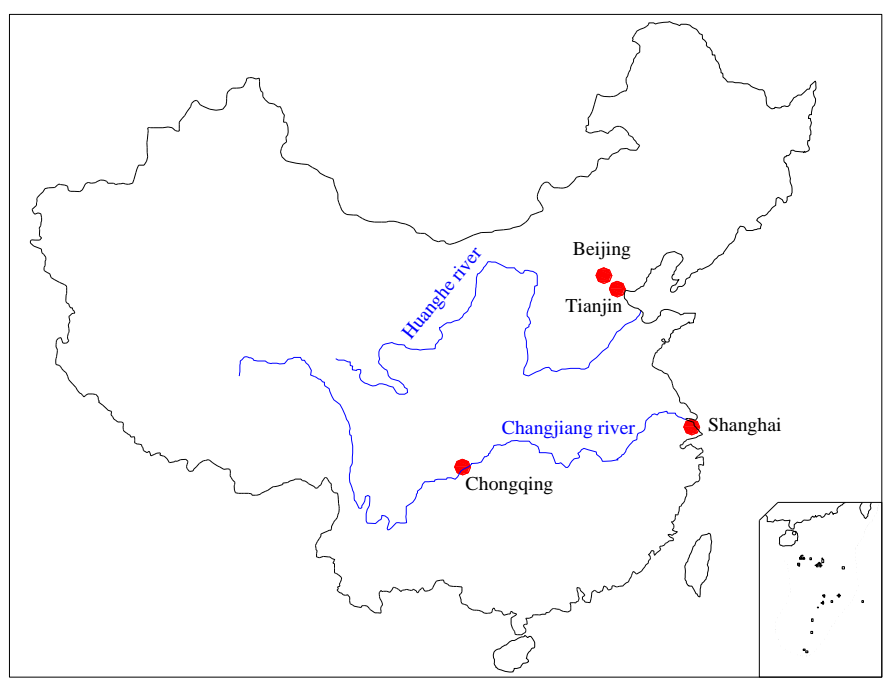

**Figure 1** The four municipalities directly under the Central Government of China: Beijing (BJ), Tianjin

(TJ), Shanghai (SH), and Chongqing (CQ)

The second approach is double logarithmic linear regression based on ordinary least square calculation. Take logarithm of all variables, and then make regression analysis of these variables in pairs. The regression coefficients are allometric scaling exponents. More simply, the least square method can be used to fit the power function directly to each pair of the original variables. The power exponents are just the allometric scaling exponents (Figure 2). The goodness of fit is the squared correlation coefficients ($R$). Tabulating the calculated values displays that the two different approaches lead to the same results (Table 2). Based on the results of any method, the multi-scaling analysis can be carried out for city development.

## 3.2 Results and analysis

The simple approach to estimating eigenvector and the maximum eigenvalue is the geometric average method. The process is as below. The first step is to calculate the geometric mean by row. The formula and results are

$$\hat{\mathbf{W}} = (\prod_{j=1}^{n} \alpha_{ij})^{1/n} = [0.9600 \quad 1.1181 \quad 0.8445 \quad 1.0898]^{\mathrm{T}}. \tag{31}$$

The second step is to normalize the entries in the eigenvector. That is

$$\hat{\mathbf{w}} = \hat{\mathbf{W}} / \sum_{i=1}^{n} W_i = [0.2393 \quad 0.2787 \quad 0.2105 \quad 0.2716]^{\mathrm{T}}, \tag{32}$$

which gives the numerical values of evaluating the cities. The third step is to compute the maximum eigenvalue. The formula and result are

$$\lambda_{\max} = \sum_{i=1}^{n} (\mathbf{M}^* \hat{\mathbf{w}})_i = 3.9878, \tag{33}$$

where $\mathbf{M}^*$ refers to the matrix of allometric scaling exponents, as indicated above, and $\mathbf{w}$, to the normalized eigenvector (Table 3).

**Table 2 The matrices of allometric scaling exponents and corresponding values of goodness of fit**

| City | Allometric scaling exponent (ASE) | | | | Goodness of fit (GOF) | | | |
| --- | --- | --- | --- | --- | --- | --- | --- | --- |
| | BJ | TJ | SH | CQ | BJ | TJ | SH | CQ |
| BJ | 1 | 0.8559 | 1.1352 | 0.8743 | 1 | 0.9945 | 0.9988 | 0.9892 |
| TJ | 1.1620 | 1 | 1.3183 | 1.0203 | 0.9945 | 1 | 0.9922 | 0.9922 |
| SH | 0.8799 | 0.7527 | 1 | 0.7679 | 0.9988 | 0.9922 | 1 | 0.9844 |
| CQ | 1.1314 | 0.9724 | 1.2820 | 1 | 0.9892 | 0.9922 | 0.9844 | 1 |

Table 3 The matrices of allometric scaling exponents, geometric mean, and normalized eigenvector

| City | BJ | TJ | SH | CQ | Geometric average | Eigenvector | ASI |
|---|---|---|---|---|---|---|---|
| BJ | 1 | 0.8559 | 1.1352 | 0.8743 | 0.9600 | 0.4757 | 0.2393 |
| TJ | 1.1620 | 1 | 1.3183 | 1.0203 | 1.1181 | 0.5541 | 0.2787 |
| SH | 0.8799 | 0.7527 | 1 | 0.7679 | 0.8445 | 0.4185 | 0.2105 |
| CQ | 1.1314 | 0.9724 | 1.2820 | 1 | 1.0898 | 0.5400 | 0.2716 |

The process of allometric scaling analysis is based on the following postulate: each pair of elements in an urban system follow the allometric scaling law. Due to spatial heterogeneity, which results in space-time translational asymmetry of geographical mathematical laws, we cannot guarantee that any pair of elements in a city or cities in a network always follows the allometric growth law (Figure 2). Therefore, it is necessary to make tests before implementing evaluation of city development. The allometric scaling exponent matrix bears an analogy with the pairwise comparison matrix of analytical hierarchical process (AHP) of Saaty (1999, 2008). So the test of AHP for positive reciprocal matrix consistency can be employed to carry out basic evaluation for the model. Defining a scaling consistency index (SCI) (Chen, 2008; Chen, 2017), we have

$$SCI = \frac{|\lambda_{max} - n|}{n-1} = 0.0041. \tag{34}$$

According to Saaty (2008), for $n=4$, the random consistency index (RCI) is RCI=0.9040. Thus, the scaling consistency ratio (SCR) is SCR=SCI/RCI=0.0045<<0.1. The result is satisfying and the allometric scaling exponent matrix passed the consistency test.

The second test is to make use of the matrix of goodness of fit (GOF). The average correlation coefficient is

$$R^* = [\frac{1}{n(n-1)}(\sum_{i=1}^{n} R_{ij}^2 - n)]^{1/2} = 0.9959. \tag{35}$$

Then, input the formula "=((FINV($\alpha$,1,$T$-2)/($T$-2))/(1+FINV($\alpha$,1, $T$-2)/($T$-2)))^0.5" into any cell in a sheet of Microsoft Excel, we can gain the critical value of $R^*$. For example, for significance level $a$=0.05 and degree of freedom (DOF) df=$T$-2=19, we should input the following formula "=((FINV(0.05,1,19)/19)/(1+FINV(0.05,1,19)/19))^0.5" in a cell. Pressing the enter key gives $R_c^*$=0.4329. Since $R^*$=0.9959>>0.4329, the GOF matrix pass the test.

The average value of GRP within the 21 year can be employed to describe the absolute level of economic development, while the allometric scaling indexes can be used to characterize the relative level of city development. Where absolute level is concerned, Shanghai is higher than Beijing, which is in turn higher than Tianjin, and Chongqing is at the end of the rank. Where relative level is concerned, Tianjin is the best one, the next is Chongqing. Beijing is higher than Shanghai, and Shanghai is the most lagging (Figure 3). Comparatively speaking, on the whole, Tianjin has a larger space to develop in future.

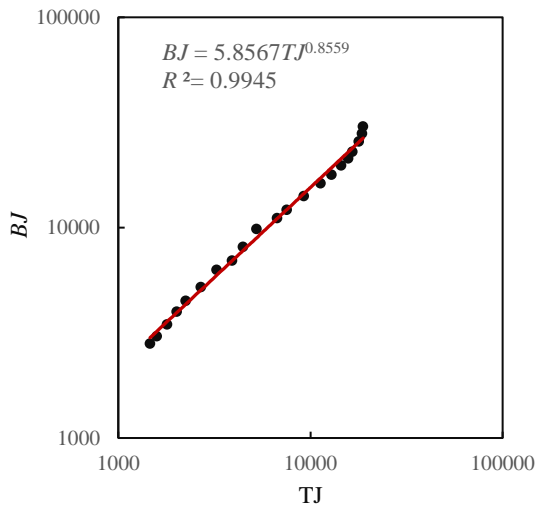

(a) Beijing vs Tianjin

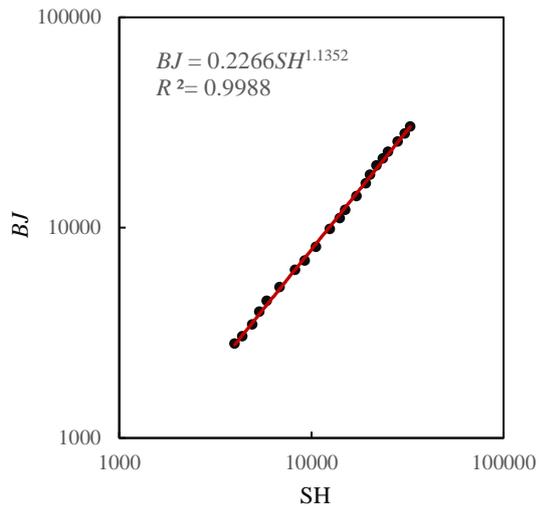

(b) Beijing vs Shanghai

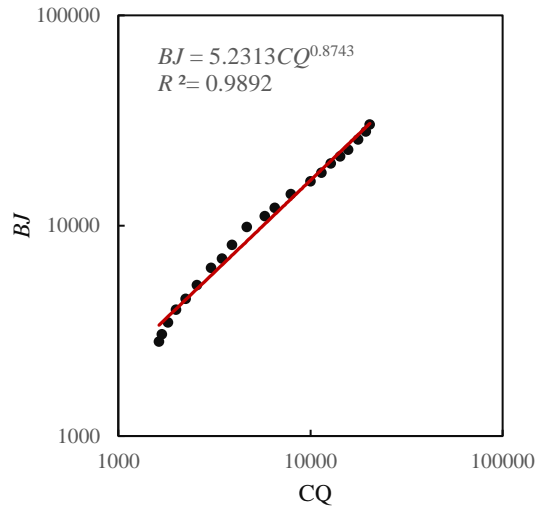

(c) Beijing vs Chongqing

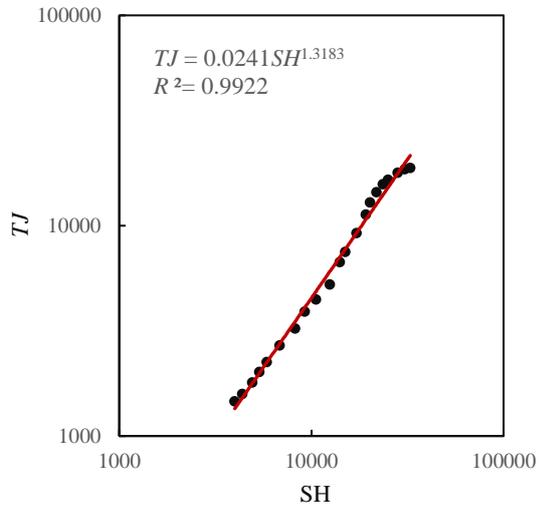

(d) Tianjin vs Shanghai

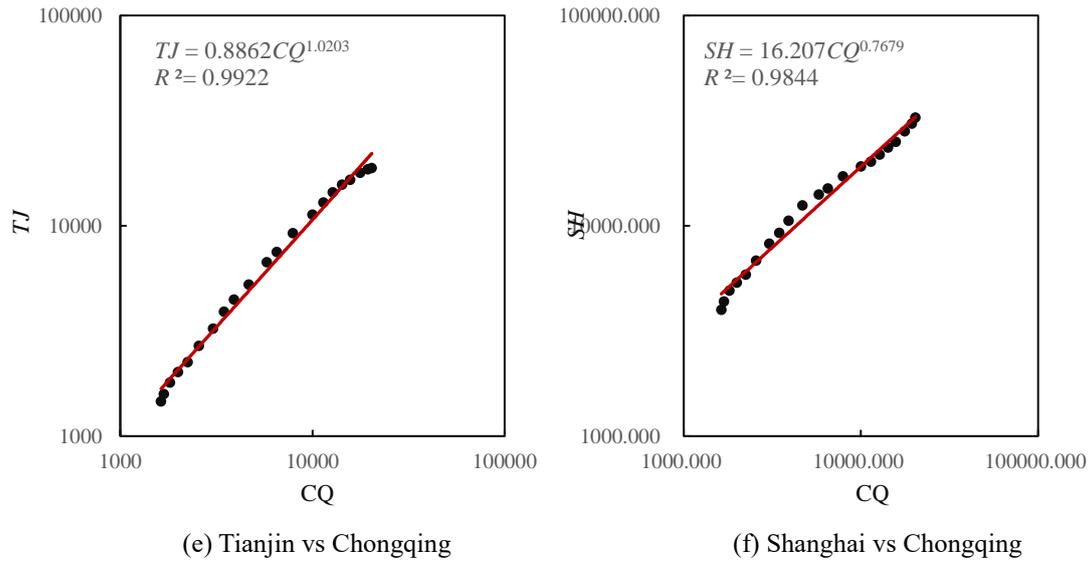

(e) Tianjin vs Chongqing  (f) Shanghai vs Chongqing

**Figure 2 The allometric scaling relations between cities of China in terms of GRP (1998-2018)**

**Note**: The numerical unit of the coordinate axes is "100 millions *yuan* (RMB)".

# 4 Discussion

The statistical essence of allometric scaling exponents has been brought to light. A rigorous mathematical demonstration was presented that an allometric scaling exponent equals the ratio of one standard deviation to another standard deviation of two correlated logarithmic variables. If the random disturbance is taken into account, an allometric scaling exponent proved to equal the product between the standard deviation ratio and correlation coefficient of two logarithmic variables. The positive analysis lends support to the theoretical derivation results. In previous works, an allometric scaling exponent proved to equal the ratio of two relative growth rates as well as the ratio of two fractal dimensions (Chen, 2008; Chen, 2017). Thus, an instructive parameter relation is derived as being expressed by equation (12). This parameter relation suggests a connection between time, space, and information. The relative growth rates reflect dynamic process, the fractal dimensions reflect spatial patterns, and the standard deviations reflect probability structure (Table 4). The three aspects represent three types of geographical space, that is, real space (common geographical space), phase space (generalized space based on time), and order space (generalized space based on hierarchy) (Chen, 2014a; Chen *et al*, 2019).

**Table 4 Three aspects of allometric scaling exponents and corresponding geographical spaces**

| Item | Measurement | Formula | Meaning | Generalized space |
|---|---|---|---|---|
| **Time** | Relative growth rate | $\alpha_{ij}=r_i/r_j$ | Dynamical process | Phase space |
| **Space** | Fractal dimension rate | $\alpha_{ij}=D_i/D_j$ | Spatial pattern | Real space |
| **Information** | Standard deviation rate | $\alpha_{ij}=\sigma_i/\sigma_j$ | Probability structure | Order space |

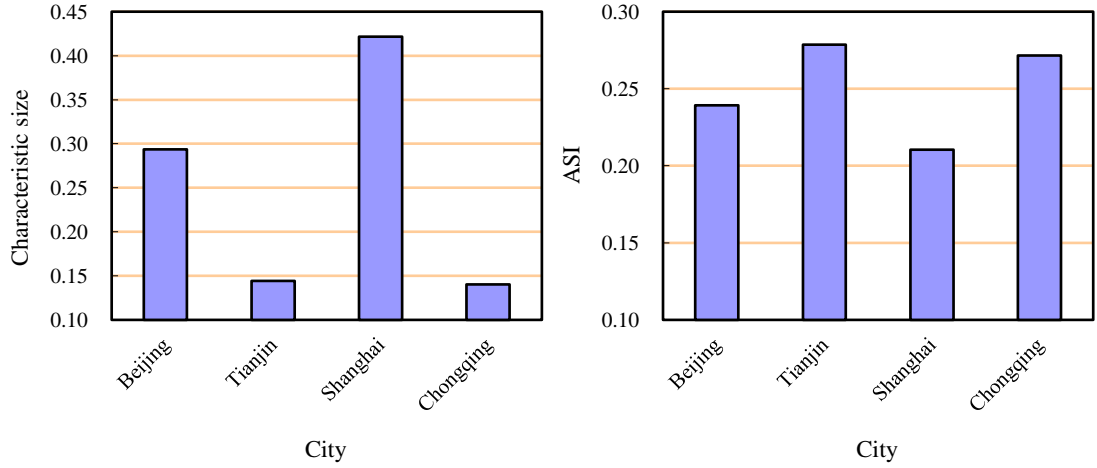

(a) Absolute development level  (b) Relative development level

**Figure 3** The absolute and relative develop levels of the four China's cities measured by average GDP and dimension scale (1998-2018)

Further, the mathematical derivation and demonstration can be generalized to fractal dimension of urban space and rank-size exponent of urban hierarchy. If we use box-counting method to estimate fractal dimension $D$, the fractal dimension value will be theoretically equal to the ratio of the standard deviation of nonempty box number logarithms $\ln N(\varepsilon)$ to the standard deviation of the linear size logarithms of boxes $\ln\varepsilon$, that is

$$D = \frac{\sqrt{\sum(\ln N(\varepsilon)-\mu_N)^2}}{\sqrt{\sum(\ln\varepsilon-\mu_\varepsilon)^2}} = \frac{\sigma_N}{\sigma_\varepsilon}, \quad (36)$$

where $\mu_N$ and $\sigma_N$ are the mean and standard deviation of $\ln N(\varepsilon)$, $\mu_\varepsilon$ and $\sigma_\varepsilon$ are the mean and standard deviation of $\ln\varepsilon$. Empirically, the fractal dimension equals the product of the ratio of two standard deviations and the correlation coefficient between box number logarithms and corresponding linear size logarithm. If we utilize Zipf's law to describe the rank-size distribution of cities, the Zipf exponent $q$ will be theoretically equal to the ratio of the standard deviation of city size logarithms $\ln P(k)$ to the standard deviation of the city rank logarithms $\ln k$, namely,

$$q = \frac{\sqrt{\sum_k (\ln P(k) - \mu_P)^2}}{\sqrt{\sum_k (\ln k - \mu_k)^2}} = \frac{\sigma_P}{\sigma_k}. \tag{37}$$

in which $\mu_P$ and $\sigma_P$ denote the mean and standard deviation of ln$P(k)$, $\mu_k$ and $\sigma_k$ refers to the mean and standard deviation of ln$k$. Empirically, Zipf's scaling exponent equals the product of the ratio of two standard deviations and corresponding correlation coefficient. In short, the standard deviation ratios provide a new way of looking at general scaling exponent from statistic viewpoint (Table 5).

**Table 5 A comparison between allometric scaling exponent, fractal dimension, and rank-size exponent**

| Item | Object | Mathematical and statistic meaning |
|---|---|---|
| **Allometric scaling exponent** | Dynamical processes | The ratio of the standard deviation of one measure logarithms to the standard deviation of another correlated measure logarithms |
| **Fractal dimension** | Spatial patterns | The ratio of the standard deviation of a measure (e.g., nonempty box numbers) logarithms to the standard deviation of corresponding scale (e.g., linear size of boxes) logarithms |
| **Rank-size exponent** | Hierarchical structure | The ratio of the standard deviation of the size (e.g., city population) logarithms to the standard deviation of corresponding rank (natural numbers) logarithms |

The multi-scaling allometric analysis bears a potential to be applied to many different fields. In nature and human society, we can find the allometric scaling relations everywhere. As indicated above, allometry proved to be a priori principle (Bertalanffy, 1968). The research of allometric growth came from the field of biology (Gayon, 2000; Gould, 1966; Lee, 1989; West, 2017). Allometric analysis was initially introduced into urban geography by Naroll and Bertalanffy (1956). Since then, a great many studies on allometric growth and allometric relations of cities have been made for a long time, and various interesting results and findings appeared successively (Batty, 2008; Batty and Longley, 1994; Chen, 2008; Lee, 1989; Lo, 2002; West, 2017). Allometric analyses fall into three categories: *temporal allometry*, *spatial allometry*, and *hierarchical allometry*. Temporal allometry is termed *longitudinal allometry* based on time series (Chen, 2014b; Naroll and Bertalanffy, 1956), spatial allometry indicates isoline allometry based on spatial distributions (Chen et al, 2019), and hierarchical allometry is equivalent to cross-sectional allometry based on rank-size distributions, which is a kind of transversal allometry (Pumain and Moriconi-Ebrard, 1997). Multi-scaling allometric analysis is based longitudinal allometry and makes use of panel data of urban growth. Maybe it can be used to integrate longitudinal allometry and transversal allometry into a systematic framework (Chen, 2008; Chen, 2017).

Allometric scaling analysis have been employed to make studies on cities based on different scales. Microscale allometry is defined at building scale or patch scale, which is involved with the allometry scaling of building geometries and the fractal measure relation between area and perimeter of urban patches (Batty *et al*, 2008; Batty and Longley, 1994; Bon, 1973; Gould, 1973). Mesoscale allometry is defined at individual city scale, which belongs to intra-urban geography. This allometry includes the allometry relations between urban area and population size based on time series (Chen, 2010; Lo and Welch, 1977), the measure relations between urban area and perimeter based on time series (Batty and Longley, 1994), and the affine relations between two orthogonal directions of urban growth (Chen and Lin, 2009). Spatial allometry is on the basis of this scale (Chen *et al*, 2019). Macroscale allometry is defined at the regional scale for systems of cities, which belongs to interurban geography. Cross-sectional data is easy to obtain, so this kind of research is the most common. A series of cases can be found as below: The allometry relations between urban area and population size based on cross-sectional data or hierarchical series (Batty and Longley, 1994; Chen, 2008; Chen, 2010; Chen and Jiang, 2018; Dutton, 1973; Lee, 1989; Lo and Welch, 1977; Nordbeck, 1971; Woldenberg, 1973), allometric relations between urban area and boundary (Batty and Longley, 1994; Benguigui *et al*, 2006; Chen and Wang, 2016), allometric relations between an urban system and its central city (Beckmann, 1958; Chen, 2008; Zhou, 1995), allometric relations between different cities of an urban system (Chen, 2008; Chen, 2017; Chen and Jiang, 2009), allometric scaling relations between urban area and road surface area as well as relations between urban area and number of bus stops in urban traffic networks (Kwon, 2018; Samaniego and Moses, 2008), and so on. In fact, the allometric scaling analysis involves various aspects of urban research, for example, allometric scaling relations between urban population, area, and various infrastructures (Arcaute *et al*; 2015; Bettencourt, 2013; Bettencourt *et al*; 2007, Bettencourt *et al*, 2010; Chen, 1995; Chen, 2008; Chen and Lin, 2009; Kühnert *et al*, 2006; Lo, 2002; Louf and Barthelemy, 2014a; Louf and Barthelemy, 2014b; Zhang and Yu, 2010), allometric scaling in urban production function (Chen, 2008; Chen and Lin, 2009, Lobo *et al*, 2013; Luo and Chen, 2014), allometric scaling relation between urban population and national total population (Dutton, 1973), allometric scaling relations between urban and rural population for urbanization (Chen, 2014; Chen *et al*, 1999; Naroll and Bertalanffy, 1956), and allometric scaling in bifurcation process of urbanization dynamics (Chen, 2009). Based on different scales, the method of multi-scaling allometric analysis can be developed to multi-level allometric analysis.

Comparing with the previous studies, the novelty of this paper lies in two respects. First, a strict mathematical proof about the relationships between allometric scaling exponents and the ratio of the standard deviation of one size measure logarithms to the standard deviation of another correlated size measure logarithms. Second, the proved result was generalized to fractal dimension and rank-size scaling exponent. Thus, the internal relationships between characteristic scales and scaling can be revealed from an angle of view. The mathematical and statistic essence of the allometric scaling exponents is helpful to understand scaling analysis and fractal characterization of cities. The main shortcomings of this study is as follows. First, the case study is too simple to show the function of multiscaling allometric analysis. The chief task of this paper is to put forward the mathematical demonstration rather than positive research. Several empirical studies have been made in previous works (Chen, 2008; Chen, 2017; Chen and Jiang, 2009; Long and Chen, 2019). Second, the multiscaling allometric analysis are based on simple allometric growth. This is an approximate treatment, which is only effective for the early stage of urban system development. The simple relations of allometric growth are derived from the first order differential equation system as follows (Bertallanffy, 1968)

$$\frac{dQ_i}{dt} = a_i Q_i, \quad \frac{dQ_j}{dt} = a_j Q_j, \tag{38}$$

where $a_i$, $a_j$ refers to the intrinsic growth coefficients, equation (38) means that the growth rate of each city depends only on the city's size itself (Chen, 2008). No direct spatial correlation between any two cities. The spatial interaction between $Q_i$ and $Q_j$ can be described with gravity model as below:

$$I_{ij} = G \frac{Q_i Q_j}{r_{ij}^b}, \tag{39}$$

where $I_{ij}$ denotes the gravity between city $i$ and city $j$, which can be represented with the quantity of the flow from one city to the other, $Q_i$ and $Q_j$ are the "mass", which can be reflected by the population size of cities $i$ and $j$, $r_{ij}$ is the distance between cities $i$ and $j$, $G$ denotes a proportionality coefficient, and $b$, refers to the distance decay exponent. Suppose that urban growth rates depend on city sizes and interaction. Integrating equation (39) into equation (38) yields

$$\begin{cases} \dfrac{dQ_i(t)}{dt} = a_i Q_i(t) + c_{ij} I_{ij} = a_i Q_i(t) + \dfrac{c_{ij} K}{r_{ij}^b} Q_i(t) Q_j(t) \\ \dfrac{dQ_j(t)}{dt} = a_j Q_j(t) + c_{ji} I_{ji} = a_j Q_j(t) + \dfrac{c_{ji} K}{r_{ij}^b} Q_j(t) Q_i(t) \end{cases}, \quad (40)$$

where $c_{ij}$, $c_{ji}$ denotes the two coupling coefficients. This is a nonlinear interaction model of two cities. For a given city pair, the distance $r_{ij}$ can be treated as a constant. Due to the coupling item $Q_i(t)Q_j(t)$, the allometric relations between and cannot be described by the simple power function, equation (1). When the city sizes based on certain measures are small, the coupling terms can be ignored. However, if the cities develop to a certain extent and the spatial interaction is strengthened, the coupling term cannot be simply omitted. The improvement of the model and related problems remain to be explored and solved in the future studies.

# 5 Conclusions

This is a theoretical study on the mathematical and statistic essence of allometric scaling exponent. The aim of this work is at demonstrating of the relations between scaling exponents and standard deviations based on logarithmic transformation. The main findings are as below. First, an allometric scaling exponent value theoretically equals the ratio of the standard deviation of one size measure logarithms to the standard deviation of another correlated size measure logarithms. This proof is based on population and time series. Second, an allometric scaling exponent value empirically equals the product between the ratio of the standard deviation of one size measure logarithms to the standard deviation of another correlated size measure logarithms and the corresponding correlation coefficient. This proof is based on samples or sample paths extracted from time series. The theoretical derivation results are verified by the empirical evidence. On the basis of these discoveries, the main conclusions can be drawn as follows. First, scaling differs from characteristic scales, but there is inherent relationships between scaling exponents and characteristic lengths. Standard deviations are based on means, and both means and standard deviations represent characteristic lengths in statistics. Allometric scaling exponents can be calculated by the ratios of two standard deviations. Maybe standard deviations depend on measurement scales, but the ratios are independent of measurement scales. Second, fractal dimension and Zipf's exponent are the standard deviation ratios based on logarithmic linear measures and sizes. By analogy, the mathematical proof about the allometric scaling exponent can be generalized to fractal dimension of urban space and rank-size scaling exponent of urban hierarchy. Fractal dimension is equal in theory to the ratio of the standard deviation of the measure logarithms to the standard deviation of corresponding linear scale logarithms, and Zipf's exponent equals theoretically the ratio of the standard deviation of the size logarithms to the standard deviation of corresponding rank logarithms. It is easy to verify the inference by mathematical experiments or observational data.

**Acknowledgements:**

This research was sponsored by the National Natural Science Foundation of China (Grant No. 41671167). The support is gratefully acknowledged.